\documentstyle[preprint,aps]{revtex}

\begin{document}
\draft
\title
{
Two-Channel Kondo Model as a Fixed Point of Local Electron-Phonon
Coupling System
}

\author
{
Hiroaki {\sc Kusunose} and Kazumasa {\sc Miyake}
}
\address{
Department of Material Physics, Faculty of Engineering Science,\\
Osaka University, Toyonaka 560
}
\maketitle

\begin{abstract}
It is shown on the basis of the multiplicative renormalization-group method of
two-loop order that the low-energy effective Hamiltonian of a strongly
coupled local
electron-phonon system is mapped to the two-channel Kondo model.
A phonon is treated as an Einstein oscillator with restricted Hilbert
space such that up to one-phonon process is taken into account.
By eliminating the high energy process of conduction electrons, it is shown
that a certain class of couplings between ion vibrations and conduction
electrons is selectively grown up.  As a result the system is reduced to the
two-channel Kondo model.
The crossover temperature $T_{\rm K}$ and the renormalized phonon frequency
$\Delta^{x}$ are expressed in terms of the mass ratio $m/M$, $m$ and $M$
being the mass of electron and ion, and the electron-phonon coupling
$g/D$, $D$ being half the bandwidth of conduction electrons.
The anomalous behaviors associated with this
renormalization can be mesuarable if the condition $T_{\rm K}>\Delta^{x}$ is
fulfilled.  It is demonstrated that such condition is satisfied when $g/D$ is
sufficiently large but in a realistic range.
\end{abstract}

\eject
\section{Introduction}
It has been recognized for these decades that a certain class of A15 compounds
have puzzling
physical properties\cite{ref:01,ref:03}.  One of the most important feature
of these compounds is the large anharmonicity of ionic oscillations,
which is exhibited, for example, as extremely small Debye-Waller factor
at zero temperature limit\cite{ref:02} and anomalously large resistivity of
the Ioffe-Regel limit at room temperature, where the mean free path of
conduction electron is comparable to the lattice constant\cite{ref:03}.
These compounds also show anomalously strong temperature dependence of
electronic properties such as the magnetic susceptibility $\chi$\cite{chi},
the Korringa constant $1/T_{1}T$ ($1/T_1$ being the nuclear relaxation
rate)\cite{korringa}, and so on
\cite{ref:03}.  These anomalies can be explained by assuming an existence of
extremely sharp peak in the density of states, of the order of
$10\sim10^{2}$ K,
around the Fermi level.  Such a sharp peak should arise from the many-body
effect because the position of the peak always stays right at the Fermi level
independently of either components of compounds or degrees of
stoichiometry\cite{ref:03}.

Similar anomalies are exhibited in ``heavy fermions" where the conduction
electrons and an array of localized spins form a coherent quasiparticle band
with extremely narrow width below the coherent temperature $T_0$.  On the
basis of this observation, the analogy between these two systems was pointed
out\cite{ref:04,ref:05}.  In A15 compounds, the origin of the ``spin" degrees
of freedom is attributed to doubly degenerate stable positions of ion in a
possible
double-well structure of ionic potential at A site of A$_3$B
compounds\cite{ref:03}.  Similar idea was proposed as a model for a
layered compound 2$H$-TaSe$_{2}$\cite{ref:10}.  Such an anharmonicity of
local-phonon can arise through strong electron-phonon
coupling\cite{ref:03,ref:04,ref:06},
which causes screening of ionic potential the electron clouds feel and in turn
decreases the restoring force for the ionic displacement leading to softening
of ion oscillations.  If the coupling is strong enough, a simple harmonic
potential for the ion may deform into a double-well\cite{ref:03,ref:06}.

In order that such system is renormalized into the Kondo model as
in heavy fermions, we have to keep the electron-ion coupling containing at
least quadratic term with respect to the ion displacement\cite{ref:04}.
This is because it is such a term that gives the hopping between the two
stable positions in the double-well corresponding to the spin-flip scattering
in the Kondo problem which is indispensable to obtain the Kondo scaling.
Along such a scenario, a crude and qualitative description of strongly coupled
electron-phonon system was given previously\cite{ref:04}.

The purpose of this paper is to investigate this problem more systematically
{\it without assuming the existense of double-well} but on the model of an
Einstein oscillator interacting with electron gas.  By restricting the
Hilbert space of phonon up to the one-phonon process, the Hamiltonian is
transformed to the form which can be treated by the formalism of Vlad\'ar and
Zawadowski\cite{ref:21}.  It is shown that the Hamiltonian would be
always renormalized into the form of Kondo model if the scaling step were not
interrupted by the {\it renormalized} level splitting of ion oscillations
which plays a role of an external pseudo-magnetic field acting on the
pseudo-spin.  In order that the ``Kondo regime" is realized, the energy scale
$T_{\rm K}$ characterizing the ``Kondo renormalization" is larger than the
renormalized level splitting $\Delta^{x}$.  Then it is shown that the
condition, $T_{\rm K}>\Delta^{x}$, is satisfied in the system with large
but realistic value of electron-ion coupling.

In the region where the Kondo model is applicable, the effective Hamiltonian
can be regarded as that of the two-channel Kondo
model\cite{ref:07,ref:08,ref:09}
because the conduction electrons have the real-spin degrees of freedom other
than that of the pseudo-spin which becomes explicit through the
Vlad\'ar-Zawadowski renormalization.  Here the real-spin degrees of freedom
play a
role of the channel there.  It has been well recognized that the two-channel
Kondo model exhibits the non-Fermi liquid fixed point in contrast with the
single-channel Kondo model.  Indeed, the two-channel model has been fully
solved by variety of
methods\cite{ref:11,ref:12,ref:13,ref:14,ref:15,ref:16,ref:17}.
Quite recently, it has been shown that the susceptibility of real-spin, or
channel, does exhibit the non-Fermi liquid behavior if the perturbations
breaking the particle-hole symmetry, such as the Coulomb repulsion among
conduction electrons or potential scattering, are taken into
account\cite{ref:18}.  This may potentially give an explanation for the
anomalous temperature dependence of the spin susceptibility $\chi$ in A15
compound V$_{3}$Si\cite{chi}.

This paper is oraganized as follows. In \S 2 we present a local phonon model
interacting with conduction electrons
and derive a simplified model describing the low-energy physics.
Next in \S 3, we derive the scaling equations following the argument by
Vlad\'ar and Zawadowski\cite{ref:21} on the basis of
the formalism of multiplicative renormalization-group\cite{ref:19,ref:20}.
In \S 4 we solve the scaling equations and determine the crossover temperature
$T_{\rm K}$ and the renormalized level splitting $\Delta^{x}$.  On this basis
we discuss the possibility of observing the anomaly associated with the
two-channel Kondo model.  In the final section, we summarize the results and
discuss their implications, especially the relation with the so-called Migdal
approximation which states the absence of the vertex correction for
electron-phonon coupling and of the renormalization of the spin
susceptiblility\cite{quinn,prange}.

\section{Model Hamiltonian}
\subsection{Local Phonon Model}
We consider an Einstein oscillator interacting with conduction electrons.
This simulates an optical phonon in A15 compounds where the transition metal
ions maintain almost the atomic nature.  In such a situation,
an electrostatic restoring force tends to be screened by the cloud of
electrons.  This increases the anharmonicity of ionic oscillations and can
make the adiabatic potential for ion displacement be even double-well like
if the electron-phonon coupling is sufficiently strong.  In order to verify
such a scenario, we start with a Hamiltonian given as follows:
\begin{eqnarray}
\label{eq:001}
&&H = H_{\rm el} + H_{\rm ph} + H_{\rm ep},\\
\label{eq:002}
&&\mbox{\hspace{2mm}}H_{\rm el} = \sum_{{\bf k},\sigma}\xi_{\bf k}
a^\dagger_{{\bf k}\sigma}a_{{\bf k}\sigma},\\
\label{eq:003}
&&\mbox{\hspace{2mm}}H_{\rm ph} = \Omega\left(b^\dagger b+\frac{1}{2}\right),\\
\label{eq:004}
&&\mbox{\hspace{2mm}}H_{\rm ep} = \sum_{{\bf kk'}\sigma}\int d{\bf r}
v({\bf r}-Q\hat{z})\exp[i({\bf k}'-{\bf k})\cdot{\bf r}]
a^\dagger_{{\bf k}\sigma}a_{{\bf k}'\sigma},
\end{eqnarray}
where $a^\dagger_{{\bf k},\sigma}$ and $b^\dagger$ denote the creation
operator for the conduction electron with the wave vector ${\bf k}$ and
the spin $\sigma$ and for the Einstein phonon with the energy $\Omega$,
respectively.  $\xi_k$ is the kinetic energy measured from the Fermi level.

The electrostatic potential $v$ in (\ref{eq:004}) is assumed to be contact
type because its range is expected to be very short of the order of
atomic radius as mentined above:
\begin{equation}
\label{eq:005}
v({\bf r}-Q\hat{z})\simeq-g\delta({\bf r}-Q\hat{z}),
\end{equation}
where $Q$ is the displacement of the ion along $z$-axis and
$g$ is a coupling constant, which is positive and of the order of a bandwidth
because its origin is the Coulomb attraction between electons and the ion.
Thus the electron-phonon interaction (\ref{eq:004}) is reduced to
\begin{equation}
\label{eq:006}
H_{\rm ep} = - g\sum_{{\bf k},{\bf k}'}\sum_{\sigma}
\exp[i(k_z' - k_z)Q]a^\dagger_{{\bf k}\sigma}a_{{\bf k}'\sigma}.
\end{equation}

\subsection{Simplification of the Model}
As it will be shown below, the interaction between ionic vibrations and
electrons near the Fermi level increases logarithmically as eliminating
high-energy processes.  Namely, the electrons in the vicinity of the Fermi
level is crucial; so that the polarization of conduction electrons, which is
expressed in terms of directional dependence of the wave vector {\bf k} near
the Fermi level, plays a crucial role.  Therefore, we first introduce the
spherical wave representation for the creation operator of conduction
electrons as follows:
\begin{equation}
\label{eq:007}
a^\dagger_{klm\sigma}=(-i)^l\frac{kR}{\sqrt{6\pi}}\int d\hat{k}
Y_{lm}(\hat{k})a^\dagger_{{\bf k},\sigma},
\end{equation}
where $Y_{lm}$ is the spherical harmonics and $R$ denotes the radius of
the system.  With the use of a linearized dispersion for conduction electrons,
$\xi_k \sim k_{\rm F}(k-k_{\rm F})/m$, $k_{\rm F}$ being the Fermi wavenumber
and $m$ the mass of conduction electron, $H_{\rm el}$ given
by (\ref{eq:002}) is reduced to
\begin{equation}
\label{eq:008}
H_{\rm el} = D\sum_{lm\sigma}\int^{1}_{-1}dk k
a^\dagger_{klm\sigma}a_{klm\sigma},
\end{equation}
where the bandwidth $D$ of conduction electrons is given as
\begin{equation}
\label{bandwidth}
D=\frac{k_{\rm F}^3R}{m\pi}.
\end{equation}

Secondly, we make a simplification of the phonon part $H_{\rm ph}$,
(\ref{eq:003}).
The displacement $Q$ and its canonical momentum $P$ are represented in terms
of the phonon operators, $b^{\dagger}$ and $b$, as follows:
\begin{eqnarray}
\label{eq:009}
&& Q = q(b+b^\dagger),\\
\label{eq:010}
&& P = \frac{1}{2qi}(b-b^\dagger),
\end{eqnarray}
where $q\equiv\sqrt{1/2M\Omega}$, $M$ being the mass of the ion.
Since the low-energy phonon states are
important, we restrict the Hilbert space of the phonon in such a way that
only the states with $n=<b^\dagger b>=0$ or $1$ are included in the low-energy
effective (fixed point) Hamiltonian.  Then, $Q$, (\ref{eq:009}), and $P$,
(\ref{eq:010}), are represented in this restricted Hilbert space as
\begin{eqnarray}
\label{eq:011}
&&Q=q\sum^{0,1}_{nn'}b^\dagger_{n}\tau^x_{nn'}b_{n'},
\mbox{\hspace{3mm}} Q^2 = q^2\sum^{0,1}_{nn'}b^\dagger_{n}
(2\delta_{nn'}-\tau^z_{nn'})b_{n'},\\
\label{eq:012}
&& P = \frac{1}{2q}\sum^{0,1}_{nn'}b^\dagger_{n}
\tau^y_{nn'}b_{n'},
\mbox{\hspace{3mm}} P^2=\frac{1}{4q^2}\sum^{0,1}_{nn'}b^\dagger_n
(2\delta_{nn'}-\tau^z_{nn'})b_{n'},
\end{eqnarray}
where $\tau^{i}$ ($i=x,y,z$) is the $i$-th component of the Pauli operator,
and $b^\dagger_n$ is the pseudo-fermion operator creating the $n$-phonon state
so that $\frac{1}{2}{\bf \tau}$ can be regarded as the pseudo-spin
corresponding to the phonon degrees of freedom.

It is noted that $(Q)^2\neq (Q^2)$ and $(P)^2\neq (P^2)$ since $(Q^2)$ and
$(P^2)$ include virtual 2-phonon process.  Indeed, the matrix elements of
$Q^{2}$ are given as follows:
\begin{eqnarray}
\label{qsquare00}
&&<0|Q^{2}|0>=<0|Q|1><1|Q|0>=q^{2},\\
\label{qsquare11}
&&<1|Q^{2}|1>=<1|Q|0><0|Q|1>+<1|Q|2><2|Q|1>={3\over 2}q^{2},\\
\label{qsquare01}
&&<1|Q^{2}|0>=<0|Q^{2}|1>=0,
\end{eqnarray}
where $|n>$ denotes the $n$-phonon state.  These matrix elements are
equivalent to the expression (\ref{eq:011}).  One can show that the same
arguments hold for the operator $P^{2}$.

For the later discussions, it is more convenient to introduce the alternative
basis for the representation of phonon states as follows:
\begin{equation}
\label{eq:013}
b^\dagger_\uparrow = \frac{1}{\sqrt{2}}(b^\dagger_0-b^\dagger_1),
\mbox{\hspace{3mm}}
b^\dagger_\downarrow=\frac{-1}{\sqrt{2}}(b^\dagger_0+b^\dagger_1).
\end{equation}
Then the non-vanishing matrix elements of $Q$ are given as
\begin{equation}
\label{eq:014}
<\uparrow|Q|\uparrow>=-q,\mbox{\hspace{3mm}}<\downarrow|Q|\downarrow>=+q,
\end{equation}
where $|\uparrow>$ denotes the state
$b^{\dagger}_{\uparrow}|$vac$>$, $|$vac$>$ being the vacuum state,
and so on.

In this representation, Eqs. (\ref{eq:011}) and (\ref{eq:012}) are
transformed as
\begin{eqnarray}
\label{eq:015}
&&Q=-q\sum^{\uparrow,\downarrow}_{\alpha\beta}b^\dagger_{\alpha}
\tau^z_{\alpha\beta}b_{\beta},
\mbox{\hspace{3mm}} Q^2 = q^2\sum^{\uparrow,\downarrow}_{\alpha\beta}
b^\dagger_{\alpha}(2\delta_{\alpha\beta}+\tau^x_{\alpha\beta})b_{\beta},\\
\label{eq:016}
&& P = \frac{1}{2q}\sum^{\uparrow,\downarrow}_{\alpha\beta}b^\dagger_{\alpha}
\tau^y_{\alpha\beta}b_{\beta},
\mbox{\hspace{3mm}} P^2=\frac{1}{4q^2}\sum^{\uparrow,\downarrow}_{\alpha\beta}
b^\dagger_\alpha(2\delta_{\alpha\beta}+\tau^x_{\alpha\beta})b_{\beta},
\end{eqnarray}
and $H_{\rm ph}$, (\ref{eq:003}), can be written in the restricted Hilbert
space as
\begin{equation}
\label{eq:017}
H_{\rm ph} = \frac{1}{2}\Omega
\sum^{\uparrow,\downarrow}_{\alpha\beta}b^\dagger_\alpha
\tau^x_{\alpha\beta}b_\beta+\Omega.
\end{equation}

Thirdly, we simplify the electron-phonon interaction $H_{\rm ep}$,
(\ref{eq:004}).  Since the exponent of
$\exp[i({\bf k}'-{\bf k})\cdot{\bf r}]$ in (\ref{eq:004}) is of the order
of $k_{\rm F}Q\sim k_{\rm F}q \sim (m/M)^{1/4}\ll 1$,
we expand the exponential with respect to $k_{\rm F}Q$ up to second order.
Then, using the expressions (\ref{eq:015}), we obtain
\begin{eqnarray}
\label{eq:018}
&&H_{\rm ep} = -g\sum_{{\bf k}{\bf k}'}\sum_{\sigma}\sum_{\alpha\beta}
a^\dagger_{{\bf k}\sigma}a_{{\bf k}'\sigma}
\left[\sum_{i}V^i_{\hat{k}\hat{k'}}b^\dagger_\alpha
\tau^i_{\alpha\beta}b_\beta+V^0_{\hat{k}\hat{k'}}\delta_{\alpha\beta}\right]
+{\cal O}((k_{\rm F}Q)^3),\\
\label{eq:019}
&&\mbox{\hspace{3mm}}V^x_{\hat{k}\hat{k'}}=-\frac{1}{2}(\hat{k_z'}-\hat{k_z})^2
k_{\rm F}^2q^2,\mbox{\hspace{3mm}}V^y_{\hat{k}\hat{k'}}=0,
\mbox{\hspace{3mm}}V^z_{\hat{k}\hat{k'}}=-i(\hat{k_z'}-\hat{k_z})k_{\rm F}q,
\\
\label{eq:020}
&&\mbox{\hspace{3mm}}V^0_{\hat{k}\hat{k'}}=1-(\hat{k_z'}-\hat{k_z})^2k_{\rm F}^2q^2.
\end{eqnarray}
It is noted that the interaction $V^x_{\hat{k}\hat{k'}}$, (\ref{eq:019}),
arises from $Q^{2}$-term and gives the pseudo-spin flip scattering which is
the heart of the Kondo effect\cite{ref:22}.

In order to treat the problem in the spherical representation, we introduce
the interaction matrices $V^i_{ll'}$ defined by
\begin{equation}
\label{eq:021}
V^i_{\hat{k}\hat{k'}}=4\pi\sum_{ll'}i^{l'-l}Y_{l0}(\hat{k})Y^{*}_{l'0}
(\hat{k'})V^i_{ll'}.
\end{equation}
It is noted that the phonon vibration along the $z$-axis interacts only with
the component of $m=0$.  It is convenient to introduce the basis of conduction
electron corresponding to those of the phonon, (\ref{eq:013}), as follows:
\begin{eqnarray}
\label{eq:022}
&&a^\dagger_{k\uparrow\sigma}=\frac{1}{\sqrt{2}}
(a^\dagger_{k00\sigma}+a^\dagger_{k10\sigma}),\\
\label{eq:023}
&&a^\dagger_{k\downarrow\sigma}=\frac{1}{\sqrt{2}}
(a^\dagger_{k00\sigma}-a^\dagger_{k10\sigma}),
\\
\label{eq:024}
&&a^\dagger_{kd\sigma}=a^\dagger_{k20\sigma}.
\end{eqnarray}
With the use of (\ref{eq:007}) and (\ref{eq:021}), the simplified Hamiltonian
is given as
\begin{eqnarray}
\label{eq:026}
&&H/D = H_{\rm el}+H_{\rm ph}+H_{\rm ep},\\
\label{eq:027}
&&\mbox{\hspace{3mm}}H_{\rm el}=\sum_{\sigma}\sum_{l}^{\uparrow,\downarrow,d}
\int^{1}_{-1}dk ka^\dagger_{kl\sigma}a_{kl\sigma},\\
\label{eq:028}
&&\mbox{\hspace{3mm}}H_{\rm ph}=\sum^{x,y,z}_{i}
\sum_{\alpha\beta}^{\uparrow,\downarrow}
\Delta^i b^\dagger_{\alpha}\tau^i_{\alpha\beta}b_{\beta},\\
\label{eq:029}
&&\mbox{\hspace{3mm}}H_{\rm ep}=\int^{1}_{-1}dk\int^{1}_{-1}dk'
\sum_{\sigma}
\sum^{\uparrow,\downarrow,d}_{ll'}\sum^{\uparrow,\downarrow}_{\alpha\beta}
\sum_{i}^{x,y,z}a^\dagger_{kl\sigma}v^i_{ll'}a_{k'l'\sigma}
b^\dagger_\alpha\tau^i_{\alpha\beta}b_\beta,
\end{eqnarray}
where $v^i_{ll'}\equiv -gV^i_{ll'}$ are the electron-phonon couplings
which are non-dimensional and $\Delta^i$ is a fictitious magnetic field
acting on the pseudo-spin $\frac{1}{2}\bf \tau$.  The explicit forms of
$v^i_{ll'}$ and $\Delta^i$ in the new basis are given as follows:
\begin{eqnarray}
\label{eq:030}
&& \hat{v}^x = \frac{1}{3}\frac{g}{D}k_{\rm F}^2q^2
\left(
\begin{array}{ccc}
0 & 1 & -1/\sqrt{10} \\
1 & 0 & -1/\sqrt{10} \\
-1/\sqrt{10} & -1/\sqrt{10} & 0
\end{array}
\right),\\
\label{eq:031}
&&\hat{v}^y= 0, \mbox{\hspace{3mm}}
\hat{v}^z = \frac{1}{\sqrt{3}}\frac{g}{D}k_{\rm F}q
\left(
\begin{array}{ccc}
1 & 0 & 0 \\
0 & -1 & 0 \\
0 & 0 & 0
\end{array}
\right),\\
\label{eq:032}
&&\Delta^x = \frac{1}{2}\frac{\Omega}{D},\mbox{\hspace{3mm}}
\Delta^y = \Delta^z = 0.
\end{eqnarray}
It is noted that the pseudo-spin ${\bf \tau}$ appears in the Hamiltonian
(\ref{eq:026}) representing the phonon degree of freedom, while it does not
in the electronic degrees of freedom at the biginning but it is induced by the
renormalization group evolution as discussed below.
It is also noted that the potential scattering is neglected in
(\ref{eq:026}), although it has been recently suggested that the potential
scattering is relevant to discuss the magnetic property on the basis of the
pseudo-spin model\cite{ref:18}.

\section{Scaling Equations}
The Hamiltonian (\ref{eq:026}) involves the logarithmic divergence at the
Fermi level due to the screening of local internal degrees of freedom by the
conduction electons as discussed by Vlad\'ar and Zawadowski\cite{ref:21}.
Since we are intersted in the physics near the Fermi level, we eliminate
the high energy process in the sense of the renormalization group evolution.
To this end, we apply the multiplicative
renormalization-group\cite{ref:19,ref:20} formalism with help of the Abrikosov
pseudo-fermion representation of the pseudo-spin\cite{ref:23}.

\subsection{Multiplicative Renormalization-Group Transformation}
First, we discuss the formalism for the multiplicative Renormalization-group
transformation on the basis of the perturbation theory.

It is useful to introduce the Matsubara Green functions for the
 electron and the pseudo-fermion defined as
\begin{equation}
\label{eq:033}
G=\frac{1}{i\omega-\xi_k},
\end{equation}
and
\begin{equation}
\label{eq:034}
{\cal G}_{\alpha\beta}=\frac{1}{i\epsilon-\lambda-\sum_{i}\Delta^i
\tau^i_{\alpha\beta}-\Sigma_{\alpha\beta}},
\end{equation}
respectively, where $\omega$ and $\epsilon$ denote the Matsubara frequency.
The electron self-energy contains a closed pseudo-fermion loop and it tends
to zero as $\lambda\rightarrow \infty$; thus, the electron self-energy is
ignored.  It is noted that ${\cal G}$ and the self-energy $\Sigma$ for
pseudo-fermion are matrices in the pseudo-spin space.

It is assumed that the system with a reduced bandwidth $D'$ behaves at low
energies in a way similar to that of original system with the bandwidth $D$,
if the couplings $v^i_{ll'}$ and the phonon energy $\Delta^i$ are modified
apropriately.  The difference between the original and the scaled Green
function and the vertex function can be given by multiplicative factors
$Z_2$ and $Z^i_{ll'}$, respectively.  The multiplicative factor
for the conduction electron is given as $Z_1=1$.
The multiplicative renormalization-group transformation can be given by
\begin{eqnarray}
\label{eq:035}
&&{\cal G}_{\alpha\beta}\left(\frac{\omega}{D'},v'^i_{ll'},\Delta'^i\right)=
Z_2\left(\frac{D'}{D},v^i_{ll'}\right){\cal G}_{\alpha\beta}
\left(\frac{\omega}{D},v^i_{ll'},\Delta^i\right),\\
\label{eq:036}
&&\gamma^i_{ll'}\left(\frac{\omega}{D'},v'^i_{ll'}\right)=
\left[Z^i_{ll'}\left(\frac{D'}{D},v^i_{ll'}\right)\right]^{-1}
\gamma^i_{ll'}\left(\frac{\omega}{D},v^i_{ll'}\right),
\end{eqnarray}
and
\begin{equation}
\label{eq:037}
v'^i_{ll'}=Z^{-1}_2Z^i_{ll'}v^i_{ll'},
\end{equation}
where $\gamma^i_{ll'}$ is the normalized vertex related to the vertex
$\Gamma^i_{ll'}$ as
\begin{equation}
\label{eq:038}
\Gamma^i_{ll'} = \gamma^i_{ll'}v^i_{ll'}.
\end{equation}
The new scaled couplings $v'^i_{ll'}$ and parameters $\Delta'^i$
are labeled by prime.
An alternative multiplicative renormalization relation for the vertex is
given by multiplying Eq. (\ref{eq:036}) by $v'^i_{ll'}$ and by inserting
the Eq. (\ref{eq:037}) and it is obtained as
\begin{equation}
\label{eq:039}
\Gamma^{i}_{ll'}\left(\frac{\omega}{D'},v^i_{ll'}\right)=Z^{-1}_2
\left(\frac{D'}{D},v^i_{ll'}\right)\Gamma^{i}_{ll'}
\left(\frac{\omega}{D},v^i_{ll'}\right).
\end{equation}
The multiplicative factor for $\Delta$ is not simple and it will be given only
along the detailed calculation.
It is important to note that the multiplicative factors depend only on the
relative change of the bandwidth $D'/D$ and on the couplings
$v^i_{ll'}$.

In order to construct the renormalization-group transformations given by
Eqs. (\ref{eq:035}) -- (\ref{eq:037}), the perturbation theory is applied.
Therefore the result to be derived is not valid in the region of strong
coupling.  According to the perturbation theory, the scaling equations is
given as
\begin{equation}
\label{eq:040}
\frac{d}{dx}v^i_{ll'} = \beta(v^i_{ll'}),
\end{equation}
and
\begin{equation}
\label{eq:041}
\frac{d}{dx}\Delta^i = f(v^i_{ll'},\Delta^i),
\end{equation}
where $x = \ln(D/D')$.
The irrelevant part of the phase space for the conduction electrons will be
eliminated by integrating Eqs. (\ref{eq:040}) and (\ref{eq:041}), but the
procedure can be applied only as far as
$x < \min[\ln(D/k_{\rm B}T), \ln(1/|{\bf \Delta'}|)]$.

\subsection{Derivation of Scaling Equations}
In order to construct the scaling equations,
it is convenient to use a new basis giving $\bar{\Delta}^z\neq 0$,
which is obtained by the rotation by the angle $\pi/2$ around the $y$-axis,
instead of the basis where $\Delta^x\neq 0$ and $\bar{\Delta}^z=0$.
In this new rotated representation, the couplings and the fictitious
magnetic field are
related as $v^x=\bar{v}^z$, $v^y=\bar{v}^y$, $v^z=-\bar{v}^x$, and
$\Delta^x=\bar{\Delta}^z$, $\Delta^y=\bar{\Delta}^y$,
$\Delta^z=-\bar{\Delta}^x$, respectively.

The vertex corrections of first and second order are shown in Fig. 1(a), 1(b)
and Fig. 2, respectively.  Their analytical expressions are given as follows:
\begin{equation}
\label{eq:042}
\Gamma^{i{\rm (I)}}_{ll'}=-2i\sum_{jk}(\bar{v}^j\bar{v}^k)_{ll'}
\epsilon^{ijk}\ln\frac{D}{|\omega|},
\end{equation}
and
\begin{equation}
\label{eq:043}
\Gamma^{i{\rm (II)}}_{ll'}=n\sum_j\left[2{\rm Tr}
(\bar{v}^i\bar{v}^j)\bar{v}^j_{ll'}
-{\rm Tr}(\bar{v}^j\bar{v}^j)\bar{v}^i_{ll'}\right]\ln\frac{D}{|\omega|},
\end{equation}
where $n$ represents a degree of freedom of spins of the conduction
electron, {\it i.e.} $n=2$.

The self-energy shown in Fig. 3 is given by
\begin{equation}
\label{eq:044}
\Sigma^{({\rm I})}_{\alpha\beta}=-n\left[\omega\sum_{i}{\rm Tr}
(\bar{v}^i\bar{v}^i)
\delta_{\alpha\beta}-\sum_{ij}\left\{2{\rm Tr}
(\bar{v}^i\bar{v}^j)\bar{\Delta}^i
-{\rm Tr}(\bar{v}^i\bar{v}^i)\bar{\Delta}^j\right\}\bar{\tau}^j_{\alpha\beta}
\right]\ln\frac{D}{|\omega|}.
\end{equation}
It is noted that the self-energy contains off-diagonal terms which are
proportional to $\bar{\tau}^x_{\alpha\beta}$ and $\bar{\tau}^y_{\alpha\beta}$.
Let us define $\bar{\Delta'}^i$ such that the renormalization factor $Z_2$ is
independent of $\bar{\Delta}^i$.   Explicit expresssions of $\bar{\Delta'}^i$
and $Z_2$ are given by
\begin{eqnarray}
\label{eq:045}
&&\bar{\Delta'}^i = \bar{\Delta}^i+2n\sum_{j}
\left[{\rm Tr}(\bar{v}^i\bar{v}^j)\bar{\Delta}^j-{\rm Tr}(\bar{v}^j\bar{v}^j)
\bar{\Delta}^i\right]\ln\frac{D}{D'},\\
\label{eq:046}
&&Z_2 = 1 + n\sum_{i}{\rm Tr}(\bar{v}^i\bar{v}^i)\ln\frac{D}{D'}.
\end{eqnarray}
It will be shown later that $\bar{\Delta}^x$ and $\bar{\Delta}^y$ can always
be eliminated by a rotation around the $y$-axis and $x$-axis in the
pseudo-spin space.
In the case $\bar{\Delta}^i=\bar{\Delta}^z\delta_{iz}$, Eqs. (\ref{eq:044}) and
(\ref{eq:045}) are rewritten as
\begin{eqnarray}
\label{eq:047}
&&\Sigma^{\rm (I)}_{\alpha\beta}=-n
\biggl[\omega\sum_{i}{\rm Tr}
(\bar{v}^i\bar{v}^i)\delta_{\alpha\beta}-2\bar{\Delta}^z{\rm Tr}
(\bar{v}^x\bar{v}^z)\bar{\tau}^x_{\alpha\beta}\nonumber\\
&&\mbox{\hspace{3mm}}-2\bar{\Delta}^z{\rm Tr}(\bar{v}^y\bar{v}^z)
\bar{\tau}^y_{\alpha\beta}
-\bar{\Delta}^z{\rm Tr}(\bar{v}^z\bar{v}^z-\bar{v}^x\bar{v}^x
-\bar{v}^y\bar{v}^y)\bar{\tau}^z_{\alpha\beta}\biggr]\ln\frac{D}{|\omega|},\\
\label{eq:048}
&&\bar{\Delta'}^i=2n\bar{\Delta}^i{\rm Tr}(\bar{v}^i\bar{v}^z)\ln\frac{D}{D'},
\mbox{\hspace{3mm}}(i = x,y),\\
\label{eq:049}
&&\bar{\Delta'}^z=\bar{\Delta}^z\left[1-2n{\rm Tr}
(\bar{v}^x\bar{v}^x+\bar{v}^y\bar{v}^y)\ln\frac{D}{D'}\right],
\end{eqnarray}
respectively.

The renormalized coupling $\bar{v'}^i_{ll'}$ is obtained by Eqs.
(\ref{eq:039}), (\ref{eq:042}), (\ref{eq:043}) and (\ref{eq:046}) as follows:
\begin{equation}
\label{eq:050}
\bar{v'}^i_{ll'}=\bar{v}^i_{ll'}-2i\sum_{jk}(\bar{v}^j\bar{v}^k)_{ll'}
\epsilon^{ijk}\ln\frac{D}{D'}-2n\sum_j\left[{\rm Tr}
(\bar{v}^j\bar{v}^j)\bar{v}^i_{ll'}-{\rm Tr}
(\bar{v}^i\bar{v}^j)\bar{v}^j_{ll'}\right]\ln\frac{D}{D'}.
\end{equation}

Although the renormalization-group transformation generates the parameters
$\bar{\Delta'}^x$ and $\bar{\Delta'}^y$, they can be eliminated by the
rotation around the $x$- and $y$-axes.
The angles of the rotations are
\begin{eqnarray}
\label{eq:051}
\alpha_x=2n{\rm Tr}(\bar{v}^y\bar{v}^z)\ln\frac{D}{D'},\\
\label{eq:052}
\alpha_y=2n{\rm Tr}(\bar{v}^x\bar{v}^z)\ln\frac{D}{D'}.
\end{eqnarray}
Then, $\bar{\Delta'}^i$ is transformed to
\begin{eqnarray}
\label{eq:055}
&&\bar{\Delta''}^x=\bar{\Delta''}^y=0,\\
\label{eq:056}
&&\bar{\Delta''}^z=\bar{\Delta}^z\left[1-2n{\rm Tr}
(\bar{v}^x\bar{v}^x+\bar{v}^y\bar{v}^y)\ln\frac{D}{D'}\right].
\end{eqnarray}
Furthermore, these rotations modify the couplings $\bar{v'}^i_{ll'}$ to
$\bar{v''}^i_{ll'}$,
\begin{eqnarray}
\label{eq:053}
&&\bar{v''}^i_{ll'}=\bar{v'}^i_{ll'}-2n{\rm Tr}(\bar{v}^i\bar{v}^z)
\bar{v}^z_{ll'}\ln\frac{D}{D'},\mbox{\hspace{3mm}}(i = x, y),\\
\label{eq:054}
&&\bar{v''}^z_{ll'}=\bar{v'}^z_{ll'}+2n\left[{\rm Tr}
(\bar{v}^x\bar{v}^z)\bar{v}^x_{ll'}+{\rm Tr}(\bar{v}^y\bar{v}^z)
\bar{v}^y_{ll'}\right]\ln\frac{D}{D'}.
\end{eqnarray}

Combining the relations (\ref{eq:050}), (\ref{eq:053}), and (\ref{eq:054}),
we obtain the scaling equations for the coupling ${\bar v}$'s as
\begin{eqnarray}
\label{eq:057}
&&\frac{d}{dx}\bar{v}^i_{ll'}=-2i\sum_{jk}\epsilon^{ijk}
(\bar{v}^j\bar{v}^k)_{ll'}
-2n\biggl[\sum_{j}\left\{{\rm Tr}(\bar{v}^j\bar{v}^j)\bar{v}^i_{ll'}
-{\rm Tr}(\bar{v}^i\bar{v}^j)\bar{v}^j_{ll'}\right\}\nonumber\\
&&\mbox{\hspace{3mm}}
+{\rm Tr}(\bar{v}^i\bar{v}^z)\bar{v}^z_{ll'}\biggr],
\mbox{\hspace{3mm}}(i=x,y),\\
\label{eq:058}
&&\frac{d}{dx}\bar{v}^z_{ll'}=-2i\sum_{jk}\epsilon^{zjk}
(\bar{v}^j\bar{v}^k)_{ll'}
-2n\biggl[\sum_{j}\left\{{\rm Tr}(\bar{v}^j\bar{v}^j)\bar{v}^z_{ll'}-
{\rm Tr}(\bar{v}^z\bar{v}^j)\bar{v}^j_{ll'}\right\}\nonumber\\
&&\mbox{\hspace{3mm}}-{\rm Tr}(\bar{v}^x\bar{v}^z)\bar{v}^x_{ll'}
-{\rm Tr}(\bar{v}^y\bar{v}^z)\bar{v}^y_{ll'}\biggr],
\end{eqnarray}
and the scaling equation for the fictitious magnetic field
${\bar {\Delta}}^{z}$ as
\begin{equation}
\label{eq:059}
\frac{d}{dx}\bar{\Delta}^z=-2n{\rm Tr}(\bar{v}^x\bar{v}^x+\bar{v}^y\bar{v}^y)
\bar{\Delta}^z.
\end{equation}

These scaling equations can be represented in the original basis of
pseudo-spin, i.e., in terms of $v$'s and $\Delta$'s without bar,
if the pseudo-spin axis is rotated back around the $y$-axis by
the angle $-\pi/2$.  Then, the scaling equations are given as follows:
\begin{eqnarray}
\label{eq:060}
&&\frac{d}{dx}v^x_{ll'}=-2i\sum_{jk}\epsilon^{xjk}
(v^jv^k)_{ll'}
-2n\biggl[\sum_{j}\left\{{\rm Tr}(v^jv^j)v^x_{ll'}-
{\rm Tr}(v^xv^j)v^j_{ll'}\right\}\nonumber\\
&&\mbox{\hspace{3mm}}-{\rm Tr}(v^xv^z)v^z_{ll'}
-{\rm Tr}(v^xv^y)v^y_{ll'}\biggr],\\
\label{eq:061}
&&\frac{d}{dx}v^i_{ll'}=-2i\sum_{jk}\epsilon^{ijk}
(v^jv^k)_{ll'}
-2n\biggl[\sum_{j}\left\{{\rm Tr}(v^jv^j)v^i_{ll'}
-{\rm Tr}(v^iv^j)v^j_{ll'}\right\}\nonumber\\
&&\mbox{\hspace{3mm}}
+{\rm Tr}(v^iv^x)v^x_{ll'}\biggr],
\mbox{\hspace{3mm}}(i=y,z),\\
\label{eq:062}
&&\frac{d}{dx}\Delta^x=-2n{\rm Tr}(v^yv^y+v^zv^z)
\Delta^x.
\end{eqnarray}
\section{Renormalization-Group Evolutions}
In this section, solving Eqs.\ (\ref{eq:060})--(\ref{eq:062}), we determine
the characteristic temperature $T_{\rm K}$, which characterizes the crossover
between weak and strong coupling regime, and the renormalized first excited
energy of phonon $\Delta^x(x)$, below which the the
renormalization-group transformation cannot be proceeded further.
Then, the typical cases of interest will be discussed.

\subsection{Solution of Scaling Equations}
First, we discuss the case $|v^x_{ll'}|$, $|v^y_{ll'}|\ll v^z_{ll'}$,
which appears at the initial stage of renormalization-group evolution of the
present problem.
In this case, we can linearize the scaling equations (\ref{eq:060}) and
(\ref{eq:061}) with respect to  $v^x_{ll'}$ and $v^y_{ll'}$.  Then, the
linearized version of the scaling equations, (\ref{eq:060}) and
(\ref{eq:061}), are given as follows:
\begin{eqnarray}
\label{eq:063}
&&\frac{d}{dx}v^x_{ll'}=-2i[v^y,v^z]_{ll'}-2n{\rm Tr}(v^zv^z)v^x_{ll'}
+4n{\rm Tr}(v^xv^z)v^z_{ll'},\\
\label{eq:064}
&&\frac{d}{dx}v^y_{ll'}=-2i[v^z,v^x]_{ll'}-2n{\rm Tr}(v^zv^z)v^y_{ll'}
+2n{\rm Tr}(v^yv^z)v^z_{ll'},\\
\label{eq:065}
&&\frac{d}{dx}v^z_{ll'}=0.
\end{eqnarray}
These equations are valid in the region of $x$ where
$v^x_{ll'}(x), v^y_{ll'}(x)\ll v^z_{ll'}(0)$.

We have verified by numerical calculations that the last
term in Eqs.\ (\ref{eq:063}) and (\ref{eq:064}) are negligible\cite{ref:24},
although we do not present it here.  Then, if we choose the representation
where $v^z_{ll'}$ is diagonal, {\it i.e.} $v^z_{ll'}=v^z_l\delta_{ll'}$,
the scaling equations above can be solved separately with the boundary
condition $v^y_{ll'}(0)=0$ in the following forms
\begin{eqnarray}
\label{eq:066}
&&v^x_{ll'}(x)=v^x_{ll'}(0)\cosh\bigl(2[v^z_{l'}(0)-v^z_l(0)]x\bigr)
\exp\bigl(-2n{\rm Tr}[v^z(0)v^z(0)]x\bigr),\\
\label{eq:067}
&&v^y_{ll'}(x)=iv^x_{ll'}(0)\sinh\bigl(2[v^z_{l'}(0)-v^z_l(0)]x\bigr)
\exp\bigl(-2n{\rm Tr}[v^z(0)v^z(0)]x\bigr),\\
\label{eq:068}
&&v^z_{ll'}(x)=v^z_l(0)\delta_{ll'}.
\end{eqnarray}

This result indicates that the couplings $v^x_{ll'}$ and $v^y_{ll'}$ with the
combination of $l$ and $l'$, for which $|v^z_{l'}(0)-v^z_l(0)|$ takes the
largest value, increases most rapidly as $x$ increases by the
renormalizaton-group step.
Therefore, we are left with only the $2\times 2$ subspace in the
$3\times3$ space of matrices $v_{ll'}$'s; i.e, we can effectively describe the
polarization degrees of freeedom of conduction electrons also by the
pseudo-spin of 1/2, i.e., $l=\uparrow$ and $\downarrow$, after the
renormalization evolution is proceeded enough.
In the subspace, the solutions can be rewritten in terms of the Pauli
matix $\sigma^i_{ll'}$ for conduction electron as
$v^i_{ll'}(x)=v^i(x)\sigma^i_{ll'}$.  Using these solutions, we can confirm
that the last term in Eqs. (\ref{eq:063}) and (\ref{eq:064})
can be neglected.

It is noed that the renormalization as above arises only if we take
into account the electron-phonon (ion) coupling
$V^x_{\hat{k}\hat{k'}}$, (\ref{eq:019}), including at least quadratic
term with respect to the ion displacement\cite{ref:04} because such a term
gives the pseudo-spin-flip scattering with conduction electrons which is the
heart of Kondo renormalization.

The $2\times 2$ subspace is well-defined in a region $x > x_1$,
where $x_1$ is determined by the conditions
\begin{eqnarray}
\label{eq:069}
&&v^x(x_1)=v^y(x_1)=\frac{1}{2}v^x(0)\exp\bigl(4v^z(0)[1-nv^z(0)]x_1\bigr),\\
\label{eq:070}
&&v^z(x_1)=v^z(0).
\end{eqnarray}
In the region $x>x_1$, the pseudo-spin degrees of freedom for conduction
electrons becomes explicit as the anharmonicity of ion vibrations grows.
Thus our electron-phonon model can be mapped to an anisotropic
two-channel Kondo model.

Next, we discuss the region $v^x(x)=v^y(x)\sim v^z(x)$.
In this region, our scaling equations can be simplified as
\begin{eqnarray}
\label{eq:071}
&&\frac{d}{dx}v^x = 4v^xv^z-4n[(v^x)^2+(v^z)^2]v^x,\\
\label{eq:072}
&&\frac{d}{dx}v^z = 4(v^x)^2[1-2nv^z],\\
\label{eq:073}
&&\frac{d}{dx}\Delta^x=-4n\Delta^x[(v^x)^2+(v^z)^2].
\end{eqnarray}
As will be shown in Appendix, the solutions of these scaling equations,
(\ref{eq:071})-(\ref{eq:073}), are given by
\begin{equation}
\label{eq:074}
x = -\frac{n}{4}\ln[v^x(0)v^z(0)]
+\frac{1}{4v^z(0)}\ln\left[\frac{4v^z(0)}{v^x(0)}\right]
+\frac{1}{8v^z(0)}\ln\left[\frac{v^z(x)-v^z(0)}{v^z(x)+v^z(0)}\right],
\end{equation}
which is an implicit equation for $v^z(x)$ and $v^x(x)$ is also given
in terms of $v^z(x)$ by the following relation
\begin{equation}
\label{eq:075}
v^x(x)=\left([v^z(x)]^2-[v^z(0)]^2\frac{1-2nv^z(x)}{1-2nv^z(0)}\right)^{1/2}.
\end{equation}

The crossover temperature $T_{\rm K}$ is given so as the condition
$v^z(x)/v^z(0)\gg1$ is satisfied for $D^{\prime}<k_{\rm B}T_{\rm K}$, where
the last term in Eq. (\ref{eq:074}) can be ignored.  Then solving
Eq.\ (\ref{eq:074}) without the last term, we obtain $T_{\rm K}$ as follows:
\begin{equation}
\label{eq:076}
\bar{T_{\rm K}}\equiv \frac{k_{\rm B}T_{\rm K}}{D}=\left(\frac{v^x(0)}{4v^z(0)}
\right)^{1/4v^z(0)}[v^x(0)v^z(0)]^{n/4}.
\end{equation}
In the crossover region, the second order terms of the
$\beta$-function, i.e., the right-hand side of the scaling equations,
become comparable to that of first order.  Thus the higher order terms of
renormalization must be taken into account for lower energy process
in gerneral.  Namely,
the perturbational renormalization-group theory can not be applied below
$T_{\rm K}$.  However, it has been known that the two-channel Kondo model
has the non-trivial fixed-point which is located in the intermediate coupling
regime given as
\begin{equation}
\label{eq:077}
{v^x}^*={v^y}^*={v^z}^*=\frac{1}{2n}.
\end{equation}
So, $v^{z}$ does not diverge even below $T_{\rm K}$ in contrast with the
conventional single channel Kondo problem.  In this sense, $T_{\rm K}$ gives
the energy scale where the crossover from weak to intermediate coupling
occurs.
Around the fixed point, (\ref{eq:077}), the effectve Hamiltonian describing
the low-energy physics is expected to be given by that of an isotropic
two-channel Kondo model.

\subsection{Renormalization of Phonon Excitation}
The first-excited energy of ion vibrations, $\Delta^x$, is renormalized by the
scaling equation (\ref{eq:062}): $\Delta^x$ softens as the renormalization
process is proceeded because ${\rm Tr}(v^{y}v^{y}+v^{z}v^{z})>0$.
This arises from the self-energy (\ref{eq:047}) and represents the physical
process of screening of spring constant of ion vibration by the conduction
electrons.  Such an effect causes the anharmonicity of the ion vibrations
logarithmically so long as the condition $x<\min[\ln(D/k_{\rm B}T),
\ln(1/\bar{\Delta}^z(x))]$ is not broken,
even if the bare couplings are weak.

The softening of $\Delta^{x}$ is caused mainly by the coupling with
$v^{z}v^{z}$ in Eq.\ (\ref{eq:062}), because $v^{z}$ is the largest coupling
through out the renormalization steps.  So, we estimate it using the scaling
equation (\ref{eq:073}) without the term $(v^x)^2$, i.e.,
\begin{equation}
\label{eq:078}
\frac{d}{dx}\ln(\Delta^x) = -4n(v^z)^2.
\end{equation}
As will be shown in Appendix, the solution of this equation is given by
\begin{equation}
\label{eq:079}
\Delta^x(x)=\Delta^x(0)[1-2nv^z(x)]^{1/2}\left(\frac{v^x(0)}{2v^z(0)}\right)
^{nv^z(0)}.
\end{equation}
Since the energy $\Delta^x(x)$ is required to exchange, or flip, the
pseudo-spin dynamically, it looks as if the pseudo-spin associated with
phonon degrees of freedom is quenched for the scattering with the conduction
electrons with low-energy scale below $\Delta^x(x)$.  In other words, the
renormalization evolutions given by Eqs.\ (\ref{eq:060})-(\ref{eq:062}) are
stopped there.

\subsection{Scaling for Typical Cases}
Now we discuss whether or not the scaling properties, discussed in the
preceding sections, are really measurable as a crossover phenomenon.

The bare couplings and phonon energy are expressed in terms of the original
parameters specifying our model as follows:
\begin{eqnarray}
\label{eq:080}
&&a\equiv\frac{1}{\sqrt{3}}k_{\rm F}q=\frac{1}{\sqrt{3}}
\sqrt{\frac{E_F}{\Omega}\frac{m}{M}}\sim\left(\frac{m}{M}\right)^{1/4},\\
\label{eq:081}
&&\mbox{\hspace{3mm}}\frac{\Omega}{D}\sim\frac{\Omega}{E_F}\sim
\left(\frac{m}{M}\right)^{1/2}\sim a^2,\\
\label{eq:082}
&&b\equiv\frac{g}{D},
\end{eqnarray}
where $E_{\rm F}$ is the Fermi energy of conduction electrons.
Typical values of parameters are $a \sim 10^{-2}$--$10^{-1}$ and
$b \sim 10^{0}$.
The crossover temperature $\bar{T_{\rm K}}$, (\ref{eq:076}), and the
first-excited energy $\Delta^x(T_{\rm K})$, (\ref{eq:079}), of ion vibrations
in the crossover region  are expressed in terms of these parameters as
\begin{eqnarray}
\label{eq:083}
&&\bar{T_{\rm K}}=2^{-\frac{1}{2ab}}a^{\frac{3nab+1}{4ab}}
b^{\frac{n}{2}},\\
\label{eq:084}
&&\Delta^x(T_{\rm K})=2^{-(nab+1)}a^{nab+2},
\end{eqnarray}
where we have assumed $v^z(T_{\rm K})\ll 1$ in Eq. (\ref{eq:079}).

$\bar{T_{\rm K}}$ and $\Delta^x(T_{\rm K})$ depend sensitively on $a$ and $b$,
and their dependences are shown in Fig. 4 and 5, respectively.
An anomalous behavior of physical quantities are observable for
sufficiently large $T_{\rm K}$ and small $\Delta^x(T_{\rm K})$, which are
realized for appropriate value of $m/M$ and large one of $g/D$.
It is because small value of $m/M$ gives extremely small $T_{\rm K}$, while
large value of $m/M$ gives large $\Delta^x(T_{\rm K})$ and small $g/D$ provides
small $T_{\rm K}$ and large $\Delta^x(T_{\rm K})$.
In other words, it is preferred that the electron-phonon coupling is large
and the bandwidth is small.  Such situation is expected to occur in d-band
metals with narrow bandwidth such as A15 and C15
compounds\cite{ref:03,varma}.

It is important to note that the region $k_{\rm B}T/D<\Delta^x(x)$ is
not reached by the renormalization-group evolution if the parameters provide
us with the crossover temperature $\bar{T_{\rm K}}<\Delta^x(x)$.
It is because the renormalization giving divergence of coupling constants
$v$'s are cut by the fictitious magnetic field acting on the pseudo-spin
at around $D^{\prime}/D=\Delta^x(x)$.

Typical cases of scaling are shown in Fig. 6.
The case of $\bar{T_{\rm K}}\gg \Delta^x(x)$ is shown in Fig. 6(a), where
$k_{\rm B}T/D$, the dotted line, is always larger than $\Delta^x(x)$,
the solid line, in the right side of figure.
In this case, anomalous behavior can be observed.
It is noted that the coupling $v^x$ and $v^y$ increase rapidly below
$T<T_{\rm K}$ while the coupling $v^z$ decreases reflecting the nature
of two-channel Kondo effect.
In the right side of figures (b) and (c), the energy scale quenching
the pseudo-spin $\Delta$ determined by the intersection between the solid
line, $\Delta^x(x)$, and the dotted line, $k_{\rm B}T/D$, gives us the
energy scale where the renormalization-group evolution should be stopped.
The case of $\bar{T_{\rm K}}<\Delta$ is shown in Fig. 6(b). The first-excited
energy of ion vibration is quenched before anomalous behavior sets in.
The case of $\bar{T_{\rm K}}\ll\Delta$ is shown in Fig. 6(c). Any anomaly
cannot be measurable because of extreamly small $T_{\rm K}$.

\section{Conclusions and Discussions}
We have discussed the structure of renormalization-group evolutions of
the strongly coupled electron-phonon system.  The phonon degrees of freedom
have been described by the Einstein oscillator.  In order to investigate the
low-energy physics, the Hilbert space of phonon has been restricted so that
the states with more than two phonons are prohibited though such states
determine the algebra of the restricted phonon operators through the
intermidiate or virtual states; and the conduction electrons are
represented by spherical harmonics of {\bf k} vectors on the Fermi surface
because their relevant degrees of freedom is those around the Fermi level.

The simplified Hamiltonian has the pseudo-spin degrees of freedom from the
beginning due to phonons in the restricted Hilbert space, while the electrons
have only a latent feature of the pseudo-spin which manifests itself
through the renormalization steps.  Namely, the scaling equations, which is
derived by the multiplicative renormalization-group method at the two-loop
level, shows that the polarization of conduction electrons corresponding to
those of the phonon are selectively grown at the initial stage of
renormalization-gruop evolution.  Thus, the effctive Hamiltonian for
low-energy physics are reduced to the anisotropic two-channel Kondo model
with renormalized fictitious magnetic field acting on the pseudo-spin.

In the anisotropic two-channel Kondo model with magnetic field, the scaling
equations provide us with two characteristic energy scales,
the crossover temperature $T_{\rm K}$ between the weak and strong coupling
region and the renormalized first-excited energy $\Delta^{x}$ of phonons
which is the fictitious magnetic filed acting on the pseudo-spin.
The anomaly associated with two-channel Kondo effect is observable when
$T_{\rm K}>\Delta^{x}$ which is realized if the mass ratio $M/m$
and the electron-phonon coupling $g/D$ are large enough.
It is expected that such a sutiation is realized in the transition
compounds with narrow bandwidth including A15 and C15 compounds.

The above result may be interpreted in terms of the adiabatic potential
the ion feels as follows.  As discussed in Refs.\ 2), 9), and 6), strong
enough electron-phonon coupling makes the harmonic potential of
individual Einstein oscillator be flattened and finally double-well shape
in general.  Namely, the adiabatic potential is deformed as shown
schematically in Fig. 7.  As a result, the first-excited energy of
ion vibrations is softed considerably due to the two-channel Kondo
correlation of pseudo spins.

A salient feature of the above result is that the electron-phonon coupling
suffers considerable renormalization due to pseudo-Kondo effect.  Namely,
the so-called Migdal approximation is not valid in this case because the
higher order terms give logarithmically divergent contributions for the
vertex correction.  This latter situation may have been overlooked
in the argument of proving a validity of the Migdal approximation.
So the way that the Migdal
approximation is broken down in the present theory appears to be somewhat
different from that discussed by Varma\cite{varma}, and Suzuki and
Motizuki\cite{suzuki} in which the fact that the correlation length of
CDW transition is very short is the origin of its breakdown.

It is well recognized that the spin susceptibility is not renormalized
by the electron-phonon interaction as far as the Migdal approximation is
valid\cite{quinn,prange,varma}.  However, it seems still remain as a
controversy whether the breakdown of the Migdal
approximation directly implies the existence of renormalization of the
spin susceptibility due to electron-phonon
coupling\cite{ref:06,varma,suzuki}.  The present result suggests that
only the electron-phonon coupling cannot afford to renormalize the
spin susceptibility which corresponds to the channel susceptibility
in our fixed-point two-channel Kondo model.  In order to obtain the
enhancement of channel susceptiblity in that model, we need
the perturbations breaking the particle-hole symmetry, such as the
Coulomb repulsion among conduction electrons or potential scattering,
as shown in Ref.\ 21) by the method of numerical renormalization group.
In this sense, together with the Coulomb repulsion among conduction electrons,
the interaction (\ref{eq:020}), representing the potential
scattering, may be important to discuss a possible anomaly of the (real)
spin susceptibility.
It is intersting to note the logarithmic temperature dependence of spin
susceptiblity observed in V$_3$Si\cite{chi} is consistent with the scenario
presented in this paper.

\newpage
\section*{Acknowledgements}
This work is supported by the Grant-in-Aid for Scientific Research (07640477),
and Monbusho International Scientific Programs (07044078) and (06044135), and
the Grant-in-Aid for Scientific Research on Priority Areas ``Physics of
Strongly Correlated Conductors" (06244104) of Ministry of Education, Science
and Culture, and the Research Fellowships of the Japan Society for the
Promotion of Science for Young Scientists.

\appendix
\section{}
In this Appendix the solutions of equations (\ref{eq:071})--(\ref{eq:073}) are
given exactly and their simplified forms including only the sigular terms are
derived for the case $v^x(x_1)\ll v^z(x_1)\ll 1$.

First, Eq. (\ref{eq:072}) is rearranged to eliminate $v^x$ as
\begin{equation}
\label{eq:a01}
(v^x)^2=\frac{dv^z/dx}{4(1-2nv^z)}.
\end{equation}
Then substituting this into  Eq. (\ref{eq:071}) multiplied by $v^x$,
we obtain
\begin{equation}
\label{eq:a02}
\frac{1}{4}\frac{d}{dx}\left[\frac{dv^z/dx}{(1-2nv^z)^2}\right]
=\frac{d}{dx}\left[\frac{(v^z)^2}{1-2nv^z}\right].
\end{equation}
It is easily integrated to give the relation
\begin{equation}
\label{eq:a03}
dx = \frac{1}{4}\frac{1}{(1-2nv^z)[(v^z)^2+2nC^2v^z-C^2]}dv^z,
\end{equation}
where $C$ is determined by the boundary condition at $x=x_1$ as
\begin{equation}
\label{eq:a04}
C=\left(\frac{[v^z(x_1)]^2-[v^x(x_1)]^2}{1-2nv^z(x_1)}\right)^{1/2}.
\end{equation}
By integrating (\ref{eq:a03}), we obtain $v^z(x)$ as an implicit form:
\begin{eqnarray}
\label{eq:a05}
&&x-x_1 = -\frac{n}{4}\ln\left[\frac{(1-2nv^z)^2}{(v^z)^2+2nC^2v^z-C^2}\cdot
\frac{[v^z(x_1)]^2+2nC^2v^z(x_1)-C^2}{[1-2nv^z(x_1)]^2}\right]\nonumber\\
&&\mbox{\hspace{3mm}}
+\frac{2n^2C^2+1}{8C\sqrt{n^2C^2+1}}\ln\left[
\frac{v^z+nC^2-C\sqrt{n^2C^2+1}}{v^z+nC^2+C\sqrt{n^2C^2+1}}\cdot
\frac{v^z(x_1)+nC^2+C\sqrt{n^2C^2+1}}{v^z(x_1)+nC^2-C\sqrt{n^2C^2+1}}\right].
\end{eqnarray}

For the case $v^x(x_1)\ll v^z(x_1)\sim v^z(0)\ll 1$, keeping only the singular
terms in right-hand side of Eq.\ (\ref{eq:a05}), we obtain
\begin{eqnarray}
\label{eq:a06}
x-x_1 = -\frac{n}{2}\ln[v^x(x_1)]-\frac{n}{4}\ln\left[\frac{2v^z(0)}{v^x(x_1)}
\right]
+\frac{1}{4v^z(0)}\ln\left[\frac{2v^z(0)}{v^x(x_1)}\right]
+\frac{1}{8v^z(0)}\ln\left[\frac{v^z-v^z(0)}{v^z+v^z(0)}\right].
\end{eqnarray}
We can also obtain the explict form for $x_1$ from Eq. (\ref{eq:069}) as
\begin{equation}
\label{eq:a07}
x_1=\frac{1}{4v^z(0)[1-nv^z(0)]}\ln\left[\frac{2v^x(x_1)}{v^x(0)}\right]
\sim \left[\frac{1}{4v^z(0)}+\frac{n}{4}\right]
\ln\left[\frac{2v^x(x_1)}{v^x(0)}\right].
\end{equation}
Then, eliminating $x_1$ from Eqs.\ (\ref{eq:a06}) and (\ref{eq:a07}),
we obtain Eq. (\ref{eq:074}) for $v^z(x)$, and then we also obtain
Eq. (\ref{eq:075}) for $v^x(x)$ from Eqs. (\ref{eq:a01}), (\ref{eq:a03})
and (\ref{eq:a04}).

Second, we eliminate $dx$ from Eq. (\ref{eq:078}) by using
Eq.\ (\ref{eq:a03}).  Then, the obtained equation can be integrated
with the boundary condition at $x=0$ as follows:
\begin{eqnarray}
\label{eq:a08}
&&\frac{\Delta^x(x)}{\Delta^x(0)}=\left(\frac{1-2nv^z}{1-2nv^z(0)}\right)^{1/2}
\times\nonumber\\
&&\mbox{\hspace{3mm}}\times
\left(\frac{v^z+nC^2_0+C_0\sqrt{n^2C^2_0+1}}{v^z+nC^2_0-C_0\sqrt{n^2C^2_0+1}}
\cdot
\frac{v^z(0)+nC^2_0-C_0\sqrt{n^2C^2_0+1}}{v^z(0)+nC^2_0+C_0\sqrt{n^2C^2_0+1}}
\right)^{nC_0/2\sqrt{n^2C^2_0+1}},
\end{eqnarray}
where $C_0$ is given by substituting $v^x(x_1)$ and $v^z(x_1)$ by $v^x(0)$ and
$v^z(0)$, respectively, in Eq. (\ref{eq:a04}).
For the case $v^x(0)\ll v^z(0)\ll 1$, keeping only the singular terms of
the original couplings, we obtain the Eq.\ (\ref{eq:079}).

\newpage
\begin{center}
\bf Figure Captions
\end{center}
\begin{description}
\item[]Fig. 1\\ Diagrams for vertex correction of first order.  Solid (dotted)
line stands for conduction electron (psudo-spin).
\item[]Fig. 2\\ Diagrams for vertex correction of second order.
\item[]Fig. 3\\ Diagram for Self-energy of pseudo-fermion of first order.
\item[]Fig. 4\\ $(m/M)^{1/4}$ dependence of $\bar{T_{\rm K}}$ and
$\Delta^x(T_{\rm K})$.
\item[]Fig. 5\\ $g/D$ dependence of $\bar{T_{\rm K}}$ and $\Delta^x(T_{\rm K})$.
\item[]Fig. 6\\ Typical cases of scaling; (a) anomalous behavior can be observed,
(b) the first-excited energy of ion vibration is quenched before
anomalous behavior sets in, (c) $T_{\rm K}$ is extremely small.
\item[]Fig. 7\\ Schematic shape of adiabatic potential of ion; (a) for the case
without electron-phonon coupling or the high-energy processes, and
(b) for the fixed-point Hamiltonian or the low-energy-processes.
\end{description}


\begin{thebibliography}{99}
\bibitem{ref:24}{These terms are neglected in Ref. 21 due to the assumption
$v^y, v^z \ll v^x$. This assumption, however, is not valid, thus it is not
trivial if these term can be neglected or not with no calculation.}
\end{thebibliography}
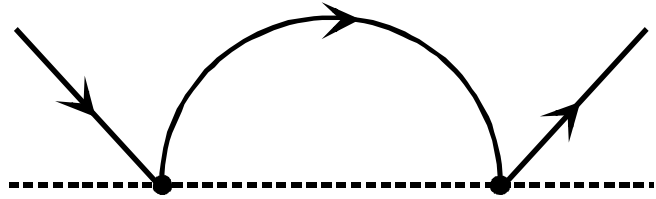
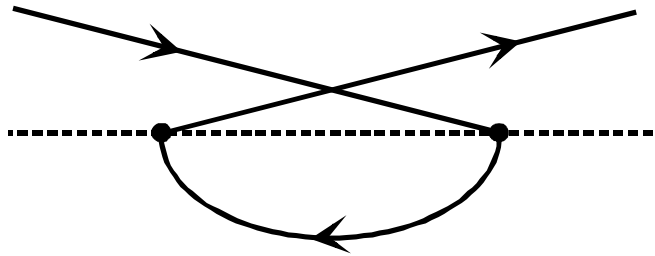
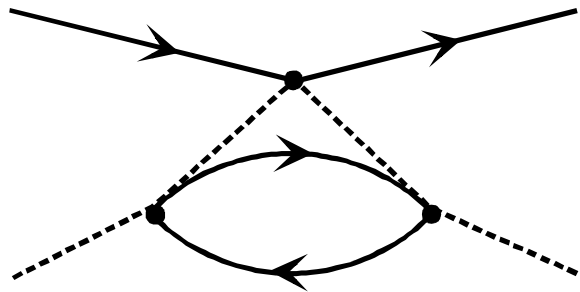
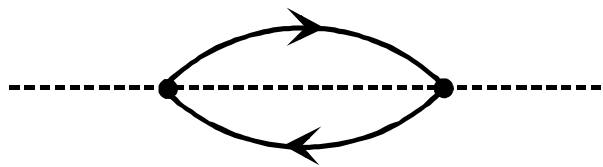
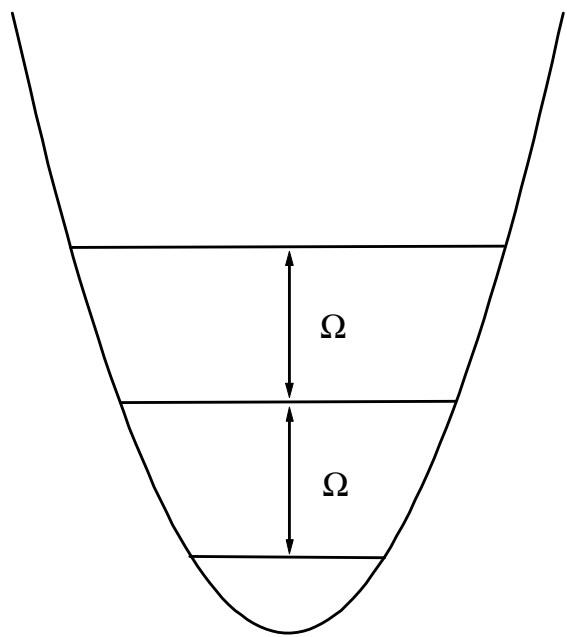
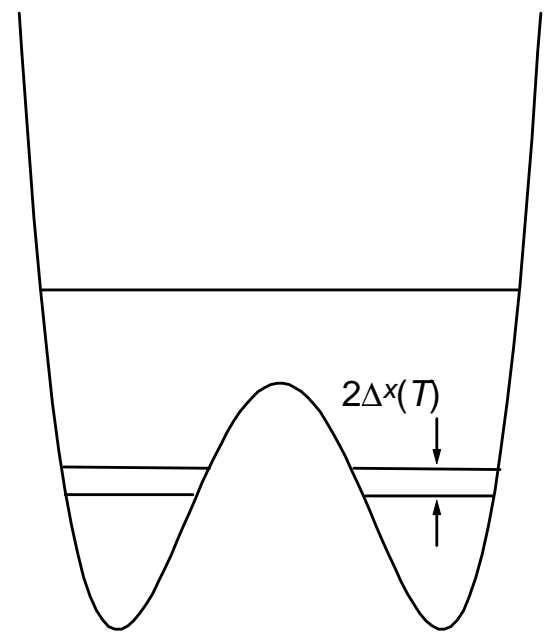
\end{document}